\documentclass{aa}
\usepackage{txfonts}
\usepackage{graphicx}
\usepackage{natbib}
\begin{document}

\title{Evidence for 9 planets in the HD 10180 system}
\author{Mikko Tuomi\thanks{\email{mikko.tuomi@utu.fi; m.tuomi@herts.ac.uk}}}
\institute{University of Hertfordshire, Centre for Astrophysics Research, Science and Technology Research Institute, College Lane, AL10 9AB, Hatfield, UK \and University of Turku, Tuorla Observatory, Department of Physics and Astronomy, V\"ais\"al\"antie 20, FI-21500, Piikki\"o, Finland}
\date{Received XX.XX.2011 / Accepted XX.XX.XXXX}

\abstract{}
{We re-analyse the HARPS radial velocities of HD 10180 and calculate the probabilities of models with differing numbers of periodic signals in the data. We test the significance of the seven signals, corresponding to seven exoplanets orbiting the star, in the Bayesian framework and perform comparisons of models with up to nine periodicities.}
{We use posterior samplings and Bayesian model probabilities in our analyses together with suitable prior probability densities and prior model probabilities to extract all the significant signals from the data and to receive reliable uncertainties for the orbital parameters of the six, possibly seven, known exoplanets in the system.}
{According to our results, there is evidence for up to nine planets orbiting HD 10180, which would make this this star a record holder in having more planets in its orbits than there are in the Solar system. We revise the uncertainties of the previously reported six planets in the system, verify the existence of the seventh signal, and announce the detection of two additional statistically significant signals in the data. If of planetary origin, these two additional signals would correspond to planets with minimum masses of 5.1$^{+3.1}_{-3.2}$ and 1.9$^{+1.6}_{-1.8}$ M$_{\oplus}$ on orbits with 67.55$^{+0.68}_{-0.88}$ and 9.655$^{+0.022}_{-0.072}$ days periods (denoted using the 99\% credibility intervals), respectively.}
{}

\keywords{Methods: Statistical, Numerical -- Techniques: Radial velocities -- Stars: Individual: HD 10180}


\maketitle


\section{Introduction}

Over the recent years, radial velocity surveys or nearby stars have provided detections of several exoplanet systems with multiple low-mass planets, even few Earth-masses, in their orbits \citep[e.g.][]{lovis2011,mayor2009,mayor2011}. These systems include a four planet system around the M-dwarf GJ 581 \citep{mayor2009}, which has been proposed to possibly have a habitable planet in its orbit \citep{vonparis2011,wordsworth2010}, a system of likely as many as seven planets orbiting HD 10180 \citep{lovis2011}, and several systems with 3-4 low mass planets, e.g. HD 20792 with minimum planetary masses of 2.7, 2.4, and 4.8 M$_{\oplus}$ \citep{pepe2011} and HD 69830 with three Neptune mass planets in its orbit \citep{lovis2006}.

Currently, one of the most accurate spectrographs used in these surveys, is the High Accuracy Radial Velocity Planet Searcher (HARPS) mounted on the ESO 3.6m telescope at La Silla, Chile \citep{mayor2003}. In this article, we re-analyse the HARPS radial velocities of HD 10180 published in \citet{lovis2011}. These measurements were reported to contain 6 strong signatures of low-mass exoplanets in orbits ranging from 5 days to roughly 2000 days and a possible seventh signal at 1.18 days. These planets include five 12 to 25 $M_{\oplus}$ planets classified in the category of Neptune-like planets, a more massive outer planet with a minimum mass of 65 M$_{\oplus}$, and a likely super-Earth with a minimum mass of 1.35 $M_{\oplus}$ orbiting the star in close proximity \citep{lovis2011}. While the confidence in the existence of the six more massive companions in this system is rather high, it is less so for the innermost super-Earth \citep{feroz2011}. Yet, even if the radial velocity signal corresponding to this low-mass companion was an artefact caused by noise and data sampling or periodic phenomena of the stellar surface, the HD 10180 system would be second only to the Solar system with respect to the number of planets in its orbits, together with the transiting Kepler-11 6-planet system \citep{lissauer2011}.

In this article, we re-analyse the radial velocity data of HD 10180 using posterior samplings and model probabilities. We perform these analyses to verify the results of \citet{lovis2011} with another data analysis method, to calculate accurate uncertainty estimates for the planetary parameters, and to see if this data set contains additional statistically significant periodic signals that could be interpreted as being of planetary origin.

\section{Observations of HD 10180 planetary system}

The G1 V star HD 10180 is a relatively nearby and bright target for radial velocity surveys with a Hipparcos parallax of 25.39$\pm$0.62 mas and $V=7.33$ \citep{lovis2011}. It is a very inactive ($\log R_{HK} = -5.00$) Solar-type star with similar mass and metallicity ($m_{\star} = 1.06 \pm 0.05$, [Fe/H]$=0.08\pm0.01$) and does not appear to show any well-defined activity cycles based on the HARPS observations \citep{lovis2011}. When announcing the discovery of the planetary system around HD 10180, \citet{lovis2011} estimated the excess variations in the HARPS radial velocities, usually referred to as stellar jitter, to be very low, approximately 1.0 ms$^{-1}$. These properties make this star a suitable target for radial velocity surveys and enable the detection of very low-mass planets in its orbit.

\citet{lovis2011} announced in 2010 that HD 10180 is host to six Neptune-mass planets in its orbit with orbital periods of 5.76, 16.36, 49.7, 123, 601, and 2200 days, respectively. In addition, they reported a 1.18 days power in the periodogram of the HARPS radial velocities that, if caused by a planet orbiting the star, would correspond to a minimum mass of only 35\% more than that of the Earth. These claims were based on 190 HARPS measurements of the variations in the stellar radial velocity between November 2003 and June 2009.

The HARPS radial velocities have a baseline of more than 2400 days, which enabled \citet{lovis2011} to constrain the orbital parameters of the outer companion in the system with almost similar orbital period. In addition, these HARPS velocities have an estimated average instrument uncertainty of 0.57 ms$^{-1}$ and a relatively good phase coverage with only seven gaps of more than 100 days, corresponding to the annual visibility cycle of the star in Chile.

\section{Statistical analyses}

We analyse the HD 10180 radial velocities using a simple model that contains $k$ Keplerian signals assumed to be caused by non-interacting planets orbiting the star. We also assume that any post-Newtonian effects are negligible in the timescale of the observations. Our statistical models are then those described in e.g. \citet{tuomi2009} and \citet{tuomi2011}, where each radial velocity measurement was assumed to be caused by the Keplerian signals, some unknown reference velocity about the data mean, and two Gaussian random variables with zero means representing the instrument noise with a known variance as reported for the HD 10180 data by \citet{lovis2011} together with the radial velocities, and an additional random variable with unknown variance that we treat as a free parameter of our model. This additional random variable describes the unknown excess noise in the data caused by the instrument and the telescope, atmospheric effects, and the stellar surface phenomena.

Clearly, the assumption that measurement noise has a Gaussian distribution might be limiting in case it was actually more centrally concentrated, had longer tails, was skewed, or was dependent on time and other possible variables, such as stellar activity levels. However, with ``only`` 190 radial velocities it is unlikely that we could spot non-Gaussian features in the data reliably. For this reason, and because as far as we know the Gaussian one is the only noise model used when analysing radial velocity data, we restrict our statistical models to Gaussian ones.

\subsection{Posterior samplings}

We analyse the radial velocities of HD 10180 using the adaptive Metropolis posterior sampling algorithm of \citet{haario2001} because it appears to be efficient in drawing a statistically representative sample from the parameter posterior density in practice \citep{tuomi2011,tuomi2011b}. This algorithm is essentialy an adaptive version of the famous Metropolis-Hastings algorithm \citep{metropolis1953,hastings1970}, that adapts the proposal density to the information gathered up to the $i$th member of the chain when proposing the $i+1$th member. It uses a Gaussian multivariate proposal density for the parameter vector $\theta$ and updates its covariance matrix $C_{i+1}$ using
\begin{equation}\label{updating}
  C_{i+1} = \frac{i-1}{i}C_{i} + \frac{s}{i} \bigg[ i \bar{\theta}_{i-1}\bar{\theta}_{i-1}^{T} - (i+1) \bar{\theta}_{i}\bar{\theta}_{i}^{T} + \theta_{i}\theta_{i}^{T} + \epsilon I \bigg] ,
\end{equation}
where $\bar{\theta}$ is the mean of the parameter vector, $( \cdot )^{T}$ is used to denote the transpose, $I$ is identity matrix of suitable dimension, $\epsilon$ is some very small number that enables the correct ergodicity properties of the resulting chain \citep{haario2001}, and parameter $s$ is a scaling parameter that can be chosen as 2.4$^{2} K^{-1}$, where $K$ is the dimension of $\theta$, to optimise the mixing properties of the chain \citep{gelman1996}.

We calculate the marginal integrals needed in model selection using the samples from posterior probability densities with the one block Metropolis-Hastings (OBMH) method of \citet{chib2001}, also discussed in \citet{clyde2007}. However, since the adaptive Metropolis algorithm is not exactly a Markovian process, only asymptotically so \citep{haario2001}, the method of \citet{chib2001} does not necessarily yield reliable results. Therefore, after a suitable burn-in period used to find the global maximum of the posterior density, during which the proposal density also converges to a multivariate Gaussian that approximates the posterior, we fix the covariance matrix to its present value, and continue the sampling with the Metropolis-Hastings algorithm, that enables the applicability of the OBMH method.

\subsection{Prior probability densities}

The prior probability densities of Keplerian models describing radial velocity data have received little attention in the literature. \citet{ford2007} proposed choosing the Jeffreys' prior for the period ($P$) of the planetary orbit, the radial velocity amplitude ($K$), and the amplitude of ``jitter``, i.e. the excess noise in the measurements ($\sigma_{J}$). This choice was justified because they make the logarithms of these parameters evenly distributed \citep{ford2007,gregory2007a}. We use this functional form of prior densities for the orbital period and choose the cutoff periods such that they correspond to the 1 day period, below which we do not expect to find any signals in this work, and a period of 10$T_{obs}$, where $T_{obs}$ is the time span of the observations. We choose this upper limit because it enables the detection of linear trends in the data corresponding to long-period companions whose orbital period cannot be constrained, but is not much greater than necessary in practice \citep[see e.g.][for the detectability limits as a function of $T_{obs}$]{tuomi2009a}, which could slow down the posterior samplings by increasing the hypervolume of the parameter space with reasonably high likelihood values. Anyhow, if the period of the outermost companion cannot be constrained from above, it would violate our detection criterion of the previous subsection.

Unlike in \citet{ford2007}, we do not use the Jeffreys' prior for the radial velocity amplitudes nor the excess noise parameter. Instead, because the HARPS data of HD 10180 deviate about their mean less than 20 ms$^{-1}$, we use uniform priors for these parameters as $\pi(K_{i}) = \pi(\sigma_{J}) = U(0, a_{RV})$, for all $i$, and use a similar uniform prior for the reference velocity ($\gamma$) about the mean as $\pi(\gamma) = U(-a_{RV}, a_{RV})$, where we choose the parameter of these priors as $a_{RV} = 20$ ms$^{-1}$. While the radial velocity amplitudes could in principle have values greater than 20 ms$^{-1}$ while the data would still not deviate more than that about the mean, we do not consider that possibility a feasible one.

Following \citet{ford2007}, we choose uniform priors for the two angular parameters in the Keplerian model, the longitude of pericentre ($\omega$) and the mean anomaly ($M_{0}$). However, we do not set the prior of orbital eccentricity ($e$) equal to a uniform one between 0 and 1. Instead, we expect high eccentricities to be less likely in this case because there are at least six, likely as many as seven, known planets ofbiting HD 10180. Therefore, high eccentricities would result in instability and therefore we do not consider their prior probabilities to be equal to the low eccentricity orbits. Our choice is then a Gaussian prior for the eccentricity, defined as $\pi(e_{i}) \propto \mathcal{N}(0, \sigma_{e}^{2})$ (with the corresponding scaling in the unit interval), where the parameter of this prior model is set as $\sigma_{e} = 0.3$. This choice penalises the high eccentricity orbits in practice but still enables them if the data insists so. In practice, with respect to the weight this prior puts on  zero eccentricity, it gives the eccentricities of 0.2, 0.4, and 0.6 relative weights of 0.80, 0.41, and 0.14, indicating that this prior can only have a relatively minor effect on the posterior densities.

Finally, we required that the planetary systems corresponding to out Keplerian solution to the data did not have orbital crossings between any of the companions. We used this condition as additional prior information by estimating that the likelihood of having any two planets in orbits that cross oneanother is zero. We could have used a more restrictive criteria, such as the requirement that the planets do not enter each others Hill spheres at any given time, but decided to keep the situation simpler because we wanted to see whether the orbital periods of the proposed companions get constrained by data instead of stability criteria, as described in the next subsection.

This choice of restricting the solutions in such a way that the corresponding planetary system does not suffer from destabilising orbital crossings also helps reducing the computational requirements by making the posterior samplings simpler. Having found $k$ Keplerian signals in the data, we simply search for additional signals between them by limiting the period space of the additional signals between these $k$ periods. We set the initial periods of the $k$ planets close to the solution of the $k$-Keplerian model and perform $k+1$ samplings where each begins with the period of the $k+1$th signal in different ''gaps`` around the previously found $k$ signals, i.e. corresponding to planet inside the orbit of the innermost one, between the two innermost ones, and so forth. If a significant $k+1$th signal is not found in one of these ''gaps``, the corresponding solution can simply be neglected. However, if there are signals in two or more gaps, it is straightforward to determine the most significant one because they can be treated as different models containing the same exact number of parameters. We then choose the most significant periodicity as the $k+1$th one and continue testing whether there are additional signals in the data. In this way, the problem of being able to rearrange the signals in any order, that would cause the posterior density to be actually highly multimodal \citep{feroz2011}, actually disappears because in a given solution the orbital crossings are forbidden and the ordering of the companions remains fixed.

\subsection{Detection threshold}

While the Bayesian model probabilities can be used reliably in assessing the relative posterior probabilities of models with differing numbers of Keplerian signals \citep[e.g.][]{ford2007,gregory2011,loredo2011,tuomi2009,tuomi2011}, we introduce additional criteria to make sure that the signals we detect can be interpreted as being of planetary origin and not arising from unmodelled features in the measurement noise or as spurious signals caused by measurement sampling. Our basic criterion is that the posterior probability of a model with $k+1$ Keplerian signals has to exceed 150 times that of a model with only $k$ Keplerian signals to claim there are $k+1$ planets orbiting the target star. We choose this threshold probability based on the considerations of \citet{kass1995}.

We require that the signals we detect in the measurements have radial velocity amplitudes, $K_{i}$ for all $i$, statistically distinguishable from zero. In practice this means, that not only the maximum \emph{a posteriori} estimate is clearly greater than zero, but that the corresponding Bayesian $\delta$ credibility sets, as defined in e.g. \citet{tuomi2009}, do not allow the amplitude to be negligible with $\delta = 0.99$, i.e. with a probability of 99\%. The second criterion is that the periods of all signals ($P_{i}$) are well defined by the posterior samples in the sense, that they can be constrained from above and below and do not get constrained purely by the condition that orbital crossings corresponding to the planetary orbits are not allowed.

To further increase the confidence of our solutions, we do not set the prior probabilities of the different models equal in our analyses. We suspect \emph{a priori}, that detecting $k+1$ planets would be less likely than detecting $k$ planets in any given system. In other words, we estimate that any set of radial velocity data would be more likely to contain $k$ Keplerian signals than $k+1$. Therefore, we set the \emph{a priori} probabilities of models with $k$ and $k+1$ Keplerian signals such that $P(k) = 2P(k+1)$ for all $k$, i.e. we penalise the model with one additional planet by a factor of two. Because of this subjective choice, if the model with $k+1$ Keplerian signals receives a posterior probability that exceeds our detection threshold of being 150 times greater than that of the corresponding model with $k$ Keplerian signals, we are likely underestimating the confidence level of the model with $k+1$ Keplerian signals relative to a uniform prior.

Physically, this choice of prior probabilities for different models corresponds to the fact that the more planets there are orbiting a star, the less stable orbits there will be left. Therefore, we estimate that if $k$ planets are being detected, there is naturally ''less room`` for an additional $k+1$th companion. However, this might be true statistically, not in an individual case, which leaves room for discussion. Yet, this and the benefit that we underestimate the significance of any detected signals encourages us to use this prior.

\subsection{Frequentist and Bayesian detection thresholds}

In addition to the Bayesian analyses described in the previous subsections, we analyse the residuals of each model using the Lomb-Scargle periodograms \citep{lomb1976,scargle1982}. As in \citet{lovis2011}, we plot the 10\%, 1\%, and 0.1\% false alarm probabilities (FAP's) to the periodograms to see the significance of the powers they contain. However, because \citet{lovis2011} calculated the FAP's in a frequentist manner by performing random permutations to the residuals while keeping the exact epochs of the data fixed and by seeing how often this random permutation produces the observed powers, we first put this periodogram approach to its philosophical context.

Generating $N$ random permutations of the residuals for each model aims at simulating a situation, where it would be possible to receive $N$ independent sets of measurements from the system of interest and seeing how often the measurement noise generates the signals corresponding to the highest powers in the periodogram. While they would not have been independent even in the case this process tries to simulate, because the exact epochs are fixed making the measurements actually dependent through the dimension of time (the measurement is actually a vector of two numbers, radial velocity and time), this approach suffers from another more significant flaw. The uncertainties of the signals removed from the data cannot be taken into account, which means that the method assumes the removed signals were known correctly. Obviously this is not the case even with the strongest signals, and even less so for the weaker ones, making the process prone to biases. Therefore, while likely producing reliable results when the signals are clear and their periods get accurate constraints, this method cannot be expected to provide reliable results in the case of extremely weak signals with large (and unknown) uncertainties. As noted by \citet{lovis2011}, when testing the significance of the 600 days signal, they could not take into account the uncertainties of the parameters of each Keplerian signal, that of the reference velocity, nor the uncertainty in estimating the excess noise in the data correctly.

The above ''frequentist`` way of performing the analyses and intepreting the consequent results is different from the Bayesian one. Because we only received one set of data, we have to base all our results on that and not some hypothetical data that would have corresponded to repetition of the original measurements. The Bayesian philosophy is to inference all the information from the data by combining it with our prior beliefs on what might be producing it. For instance, as described above when discussing our choice of priors, we can expect tightly packed multiplanet systems to be more likely to contain planets on close-circular orbits than on very eccentric ones. Also, with the powerful posterior sampling algorithms available, it is possible to take the uncertainties in every parameter into account simultaneously, which enables the detection of weak signals in the data \citep[e.g.][]{gregory2005,gregory2007a,gregory2007b,tuomi2009} and prevents the detection of false positives, as happened in the case of Gliese 581 \citep{vogt2010,gregory2011,tuomi2011}.

Yet, despite the above problems in the traditional periodogram analyses, we take advantage of the power spectra of the residuals in our posterior samplings. The highest peaks in the periodogram can be used very efficiently together with Bayesian methods by using the corresponding periodicities as initial states of the Markov chains in the adaptive Metropolis algorithm. Because of this choice, the initial parameter vector of the Markov chain starts very close to the likely maximum \emph{a posteriori} (MAP) solution, which makes its convergence to the posterior density reasonably rapid and helps reducing computational requirements.

\section{Results}

When drawing a sample from the parameter posterior density and using it to calculate the corresponding model probabilities, it became crucial that this sample was a statistically representative one. While posterior samplings generally provide a global solutions, it is always possible that the chain converges to a local maximum and stays in its vicinity within a sample of finite size. To make sure that we indeed received the global solution, we calculated several Markov chains starting from the vicinity of the apparent MAP solution and compared them to see that they indeed corresponded to the same posterior probability density. In practice, sampling the parameter spaces was computationally demanding because the probability that the parameter vectors drawn from the Gaussian proposal density are close to the posterior maximum decreases rapidly when the number of parameters with non-Gaussian probability densities increases. Therefore, while models with 0-6 planets were reasonably easy to sample and we received acceptance rates of 0.1-0.3, these rates decreased when increasing the number of signals in the model further. As a result, for models with 8-9 Keplerian signals, the acceptance rates decreased below 0.1 and forced us to increase the chain lenghts by two orders of magnitude from a typical $1.0 \times 10^{6}$ to as high as $1.0 \times 10^{8}$.

In the following subsections, the results are based on several samplings that yielded the same posterior densities, and also consistent model probabilities.

\subsection{The number of significant periodicities}

The posterior probabilities of the different models provide information on the number of Keplerian signals ($k$) in the data. \citet{lovis2011} were confident that there are six planets orbiting HD 10180 based on their periodogram-based analyses of model residuals and the corresponding random permutations of them when calculating the significance levels of the periodogram powers. They also concluded that the six planets in the system with increasing periods of 5.75962$\pm$0.00028, 16.3567$\pm$0.0043, 49.747$\pm$0.024, 122.72$\pm$0.20, 602$\pm$11, and 2248$^{+102}_{-106}$ days comprise a stable system given that the masses of the planets are within a factor of $\sim$ 3 from the minimum masses of 13.70$\pm$0.63, 11.94$\pm$0.75, 25.4$\pm$1.4, 23.6$\pm$1.7, 21.4$\pm$3.0, and 65.3$\pm$4.6 M$_{\oplus}$, respectively.

Because \citet{lovis2011} pointed out that there are actually two peaks in the periodogram corresponding to the seventh signal, i.e. 1.18 and 6.51 days periodicities that are the one day aliases of each other, but noted that the 6.51 signal, if corresponding to a planet, would cause the system to be unstable in short timescales, we adopted the 1.18 periodicity as the seventh signal in the data. The relative probabilities of the models with $k = 0, ..., 9$ are shown in Table \ref{model_probabilities} together with the period ($P_{s}$) of the next Keplerian signal added to the model.

\begin{table}
\center
\caption{The relative posterior probabilities of models with $k=0, ..., 9$ Keplerian signals ($\mathcal{M}_{k}$) given radial velocities of HD 10180 (or data, $d$) together with the periods ($P_{s}$) of the signals added to the model when increasing the number of signals in the model by one. Also shown are the logarithmic Bayesian evidences ($P(d | \mathcal{M}_{k})$) and their uncertainties as standard deviations and the root mean square (RMS) values of the residuals for each model.}\label{model_probabilities}
\begin{tabular}{lcccc}
\hline \hline
$k$ & $P(\mathcal{M}_{k} | d)$ & $\log P(d | \mathcal{M}_{k})$ & RMS [ms$^{-1}$] & $P_{s}$ [days] \\
\hline
0 & 1.5$\times10^{-125}$ & -621.29 $\pm$ 0.03 & 6.29 & -- \\
1 & 1.6$\times10^{-114}$ & -595.15 $\pm$ 0.04 & 5.39 & 5.76 \\
2 & 1.5$\times10^{-98}$ & -556.87 $\pm$ 0.02 & 4.34 & 123 \\
3 & 4.1$\times10^{-86}$ & -528.44 $\pm$ 0.07 & 3.62 & 2200 \\
4 & 2.7$\times10^{-53}$ & -452.17 $\pm$ 0.13 & 2.43 & 49.8 \\
5 & 7.2$\times10^{-20}$ & -374.42 $\pm$ 0.05 & 1.59 & 16.35 \\
6 & 3.4$\times10^{-13}$ & -358.36 $\pm$ 0.12 & 1.41 & 600 \\
7 & 3.8$\times10^{-7}$ & -343.73 $\pm$ 0.09 & 1.32 & 1.18 \\
8 & 0.003 & -334.06 $\pm$ 0.06 & 1.24 & 67.5 \\
9 & 0.997 & -327.79 $\pm$ 0.06 & 1.18 & 9.65 \\
\hline \hline
\end{tabular}
\end{table}

According to the model probabilities in Table \ref{model_probabilities}, the 8- and 9-Keplerian models are the most probable descriptions of the processes producing the data out of those considered. Improving the statistical model by adding the seventh signal, with a period of 1.18 days, increases the model probability by a factor of more than $10^{6}$, which makes the credibility of this seventh signal high. Adding two more signals corresponding to 67.6 and 9.66 days periodicities increases the model probabilities even further. As a result, the 9-Keplerian model receives the greatest posterior probability of slightly more than 150 times more than the next best model, the 8-Keplerian model. This enables us to conclude, that there is strong evidence in favour of the 67.6 and 9.66 days periodicities not being produced by random processes, i.e. measurement noise.

\subsection{Periodograms of residuals}

\begin{figure}
\center
\includegraphics[width=0.49\textwidth]{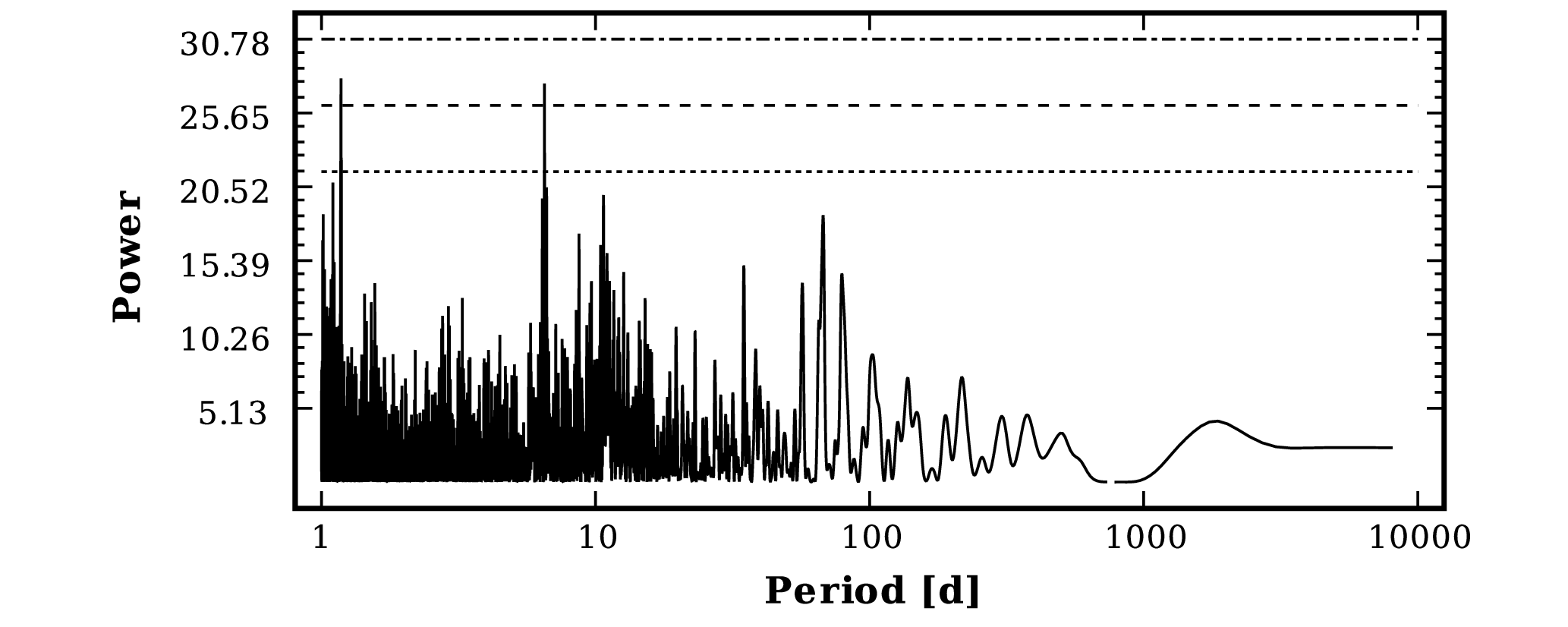}

\includegraphics[width=0.49\textwidth]{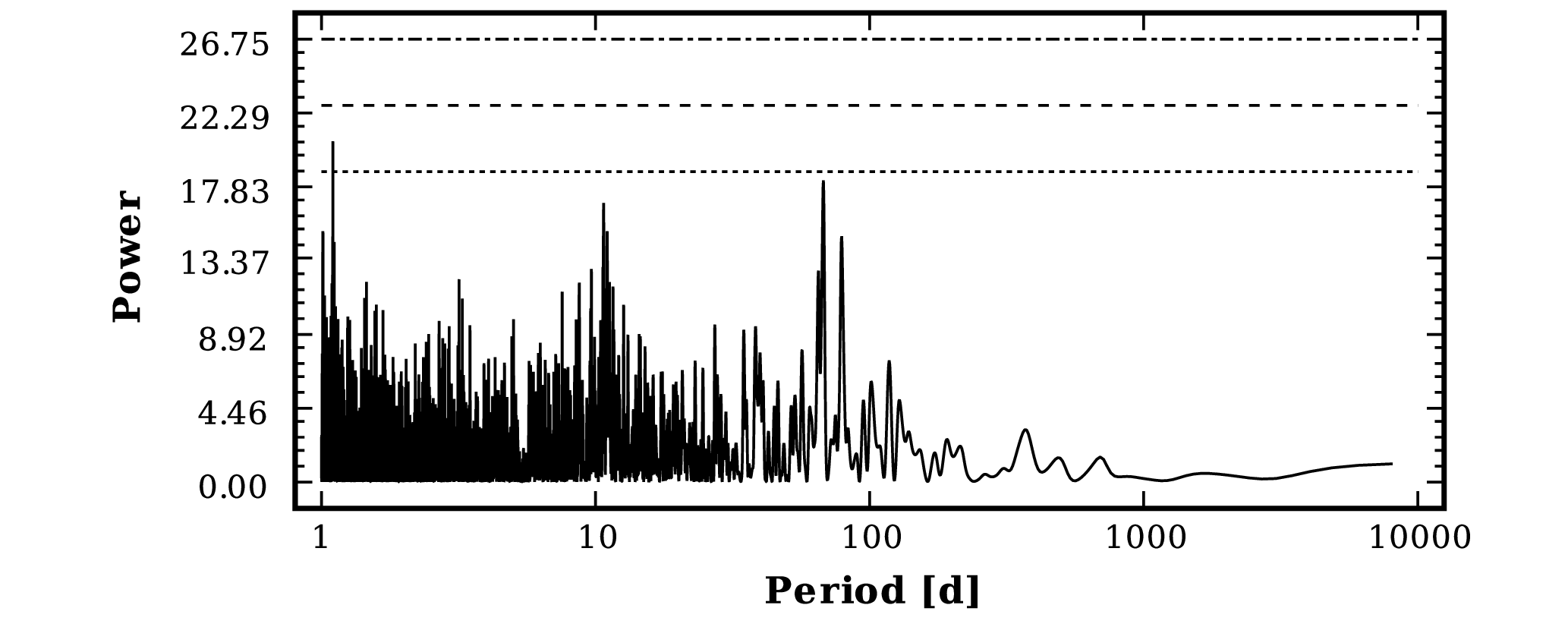}

\includegraphics[width=0.49\textwidth]{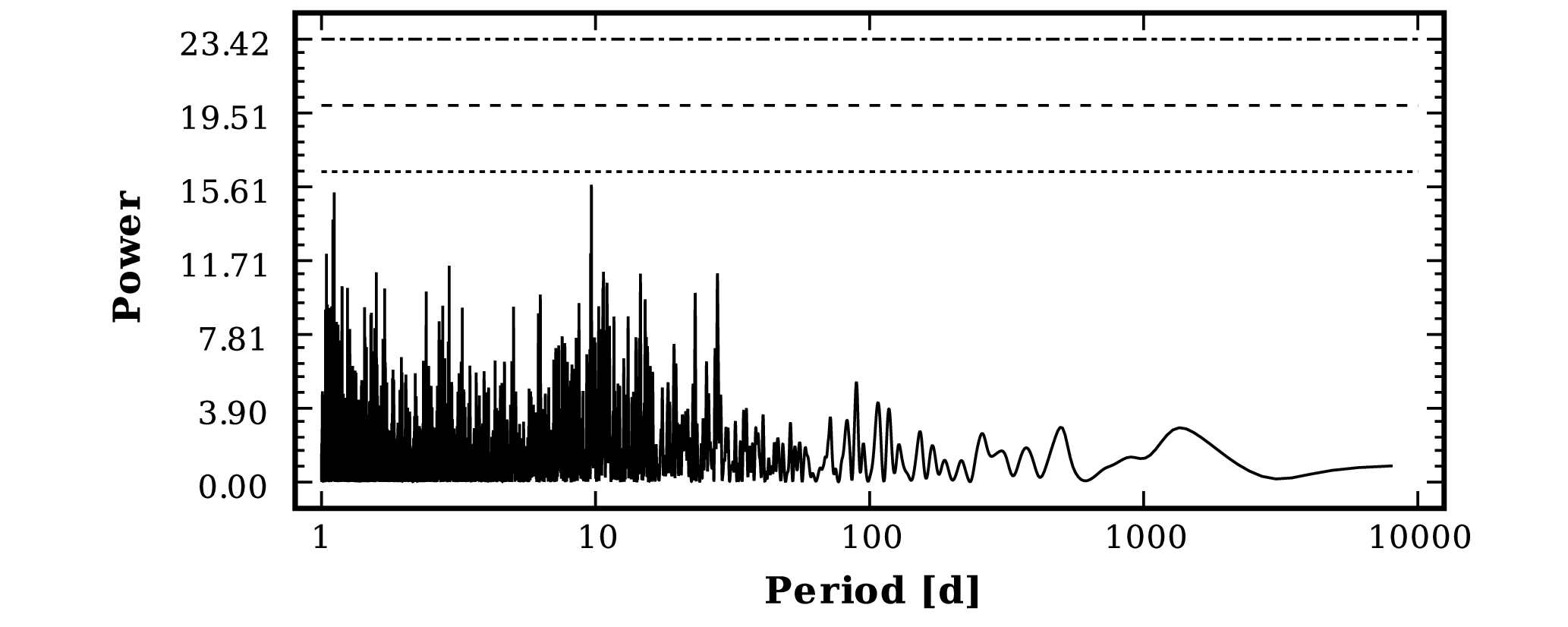}

\includegraphics[width=0.49\textwidth]{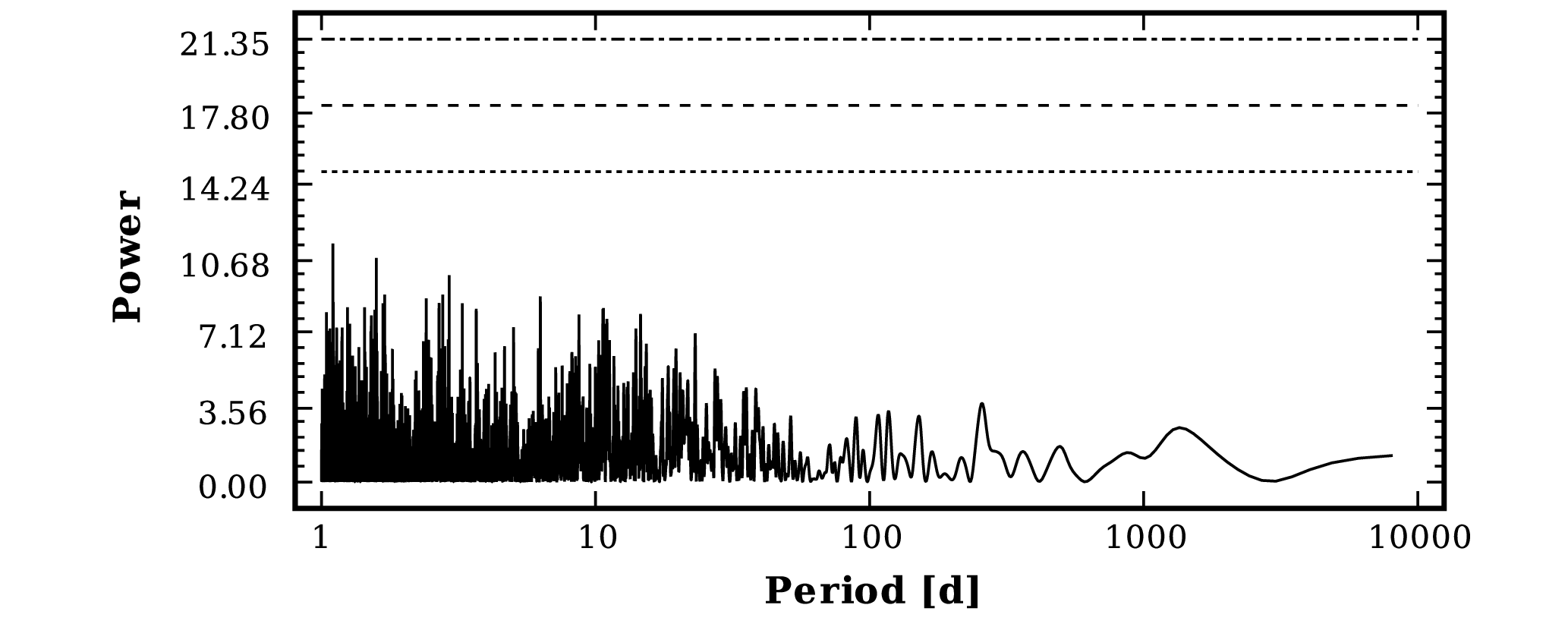}
\caption{The Lomb-Scargle periodograms of the HD 10180 radial velocities for the residuals of the models with 6 (top) to 9 (bottom) periodic signals. The dotted, dashed, and dot-dashed lines indicate the 10\%, 1\%, and 0.1\% FAPs, respectively.}\label{periodograms}
\end{figure}

We subtracted the models with 6 to 9 periodic signals from the data and calculated the Lomb-Scargle periodograms \citep{lomb1976,scargle1982} of these residuals (Fig. \ref{periodograms}) together with the standard analytica FAP's. As already seen in \citet{lovis2011}, the two strong powers corresponding to 1.18 days periodicity and its 1 day alias at 6.51 days are strong in the residuals of the 6-Keplerian model (top panel in Fig. \ref{periodograms}) and exceed the 1\% FAP. These peaks get also removed from the residuals of the 7-Keplerian model (2nd panel). However, it can be seen that 9.66 and 67.6 days periods have strong powers in these residuals, yet, neither of them exceeds even the 10\% FAP level. Modelling these periodicities as Keplerian signals and plotting the periodograms of the corresponding residuals of the 9-Keplerian model (bottom panel) shows that there are no strong powers left in the residuals. While this does not indicate that these two periodicities are significant, it shows that they are clearly present in the periodograms of the residuals and support the findings in the previous subsection based on the model probabilities.

We note that there appear to be two almost equally strong peaks in the 8-Keplerian model residuals (Fig. \ref{periodograms}, panel 3). However, these powers corresponding to periods of 9.66 and 1.11 days are one day aliases of oneanother. This aliasing is clear as the 1.11 days power is absent in the periodogram of the 9-Keplerian model (Fig. \ref{periodograms}, bottom panel).

\subsection{Planetary interpretation: orbital parameters}

Because of the samplings of parameter posterior densities of each statistical model, we were able to calculate the estimated shapes of the parameter distributions for each model and use these to estimate the features in the corresponding densities. We describe these densities using three numbers, the MAP estimates and the corresponding 99\% Bayesian credibility sets (BCS's), as defined in e.g. \citet{tuomi2009}. The simple MAP point estimates and the corresponding 99\% BCS's do not represent these dentities very well because some of the parameters are highly skewed and have tails on one or both sides. However, we list the parameters interpreted as being of planetary origin in this subsection.

When calculating the semimajor axes and minimum masses of the planets, we took the uncertainty in the stellar mass into account by treating it as an independent random variable. We assumed that this random variable had a Gaussian density with mean equal to the estimate given by \citet{lovis2011} of 1.06 M$_{\odot}$ and a standard deviation of 5\% of this estimate. As a consequence, the densities of these parameters are broader than they would be if using a fixed value for the stellar mass, which indicates greater uncertainty in their values.

\subsubsection{The 6 planet solution}

The parameter estimates of our 6-Keplerian model are listed in Table \ref{parameters6}\footnote{\citet{lovis2011} use letter b for the 1.18 days periodicity not present in the 6-Keplerian model. Therefore, we use letters c - h in Table 2 to have the letters denote the same signals. Yet, the solution of \citet{feroz2011} denotes the 5.76 days signal with letter b.}. This solution is consistent with the solution reported by \citet{lovis2011} but the uncertainties are slightly greater likely because we took into account the uncertainty in the jitter parameter $\sigma_{j}$ and because \citet{lovis2011} used more conservative uncertainty estimates from the covariance matrix of the parameters that does not take the nonlinear correlations between the parameters into account. The greatest difference is therefore in the uncertainty of the orbital period of the outermost companion, whose probability density has a long tail towards longer periods and periods as high as 2670 days cannot be ruled out with 99\% confidence (the supremum of the 99\% BCS).

\begin{table*}
\center
\caption{The 6-planet solution of HD 10180 radial velocities. MAP estimates of the parameters and their 99\% BCS's.\label{parameters6}}
\begin{tabular}{lccc}
\hline \hline
Parameter & HD 10180 c & HD 10180 d & HD 10180 e \\
\hline
$P$ [days] & 5.7596 [5.7588, 5.7606] & 16.354 [16.340, 16.368] & 49.74 [49.67, 49.82] \\
$e$ & 0.06 [0.00, 0.17] & 0.12 [0.00, 0.29] & 0.01 [0.00, 0.14] \\
$K$ [ms$^{-1}$] & 4.54 [4.06, 5.02] & 2.85 [2.35, 3.34] & 4.32 [3.74, 4.83] \\
$\omega$ [rad] & 5.8 [-] & 5.8 [-] & 2.6 [-] \\
$M_{0}$ [rad]& 4.7 [-] & 5.8 [-] & 3.7 [-] \\
$m_{p} \sin i$ [M$_{\oplus}$] & 13.2 [11.2, 15.1] & 11.8 [9.5, 14.2] & 25.6 [21.5, 29.7] \\
$a$ [AU] & 0.0641 [0.0608, 0.0673] & 0.1287 [0.1220, 0.1349] & 0.270 [0.256, 0.283] \\
\hline
& HD 10180 f & HD 10180 g & HD 10180 h \\
\hline
$P$ [days] & 122.76 [122.152, 123.44] & 603 [568, 642] & 2270 [2020, 2670] \\
$e$ & 0.06 [0.00, 0.26] & 0.05 [0.00, 0.49] & 0.03 [0.00, 0.31] \\
$K$ [ms$^{-1}$] & 2.88 [2.28, 3.42] & 1.46 [0.71, 2.21] & 2.97 [2.40, 3.68] \\
$\omega$ [rad] & 5.8 [-] & 6.0 [-] & 2.8 [-] \\
$M_{0}$ [rad]& 4.9 [-] & 4.5 [-] & 3.7 [-] \\
$m_{p} \sin i$ [M$_{\oplus}$] & 23.1 [18.2, 28.4] & 19.4 [10.2, 29.8] & 64.5 [51.5, 78.9] \\
$a$ [AU] & 0.491 [0.468, 0.518] & 1.42 [1.33, 1.52] & 3.44 [3.15, 3.88] \\
\hline
$\gamma$ [ms$^{-1}$] & -0.43 [-0.87, -0.02] \\
$\sigma_{j}$ [ms$^{-1}$] & 1.40 [1.18, 1.70] \\
\hline \hline
\end{tabular}
\end{table*}

\subsubsection{The 9 planet solution}

\begin{figure*}
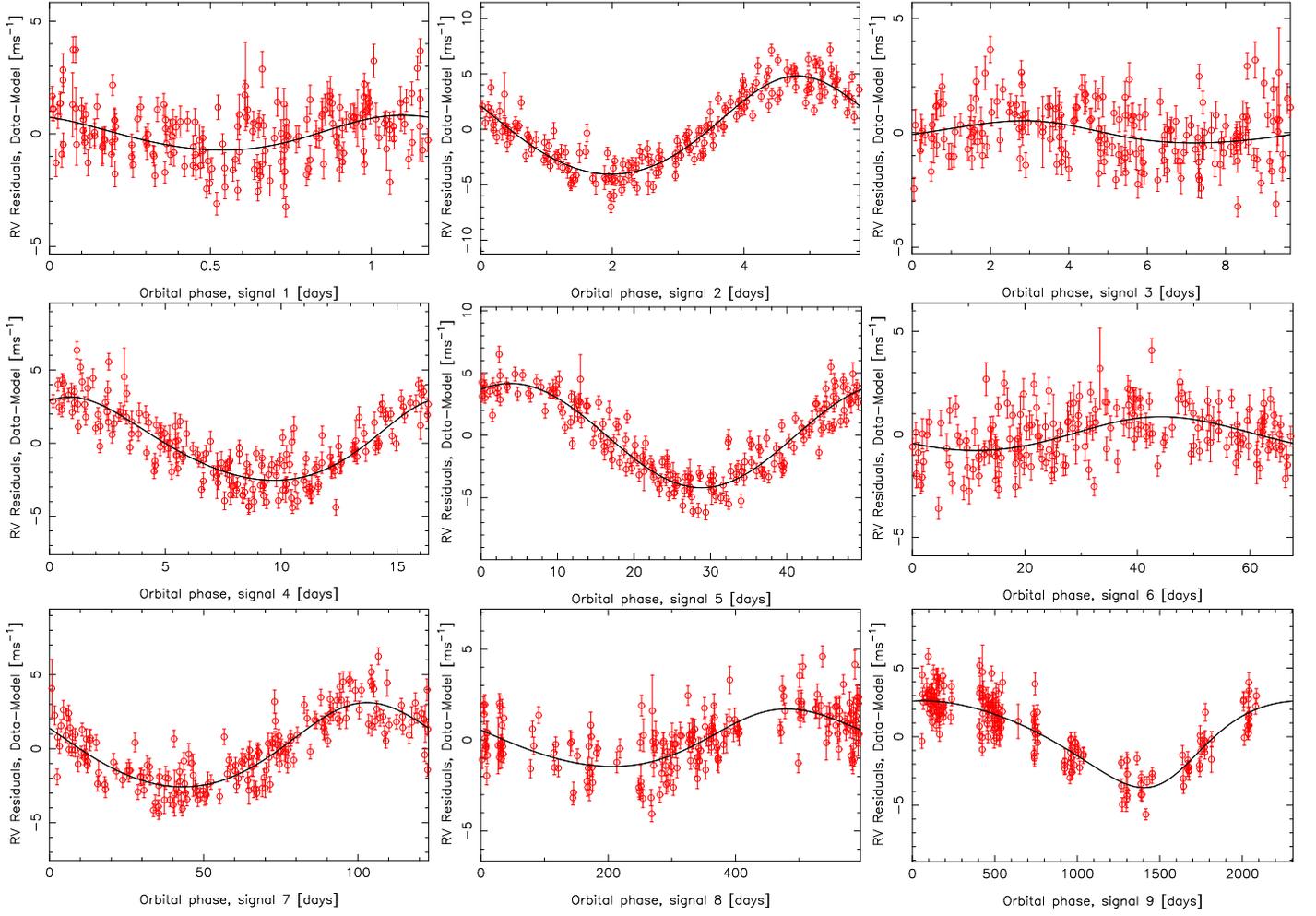

\center
\includegraphics[angle=-90,width=0.33\textwidth]{rvdist9_scresidc_rv_HD10180_1.ps}
\includegraphics[angle=-90,width=0.33\textwidth]{rvdist9_scresidc_rv_HD10180_2.ps}
\includegraphics[angle=-90,width=0.33\textwidth]{rvdist9_scresidc_rv_HD10180_3.ps}

\includegraphics[angle=-90,width=0.33\textwidth]{rvdist9_scresidc_rv_HD10180_4.ps}
\includegraphics[angle=-90,width=0.33\textwidth]{rvdist9_scresidc_rv_HD10180_5.ps}
\includegraphics[angle=-90,width=0.33\textwidth]{rvdist9_scresidc_rv_HD10180_6.ps}

\includegraphics[angle=-90,width=0.33\textwidth]{rvdist9_scresidc_rv_HD10180_7.ps}
\includegraphics[angle=-90,width=0.33\textwidth]{rvdist9_scresidc_rv_HD10180_8.ps}
\includegraphics[angle=-90,width=0.33\textwidth]{rvdist9_scresidc_rv_HD10180_9.ps}
\caption{The phase-folded Keplerian signals of the 9-planet solution with the other 8 signals removed.}\label{orbits}
\end{figure*}

Assuming that all the periodic signals in the data are indeed caused by planetary companions orbiting the star, the parameters of out 9-Keplerian solution are listed in Table \ref{parameters9} and the phase folded orbits of the 9 Keplerian signals are plotted in Fig. \ref{orbits}.

\begin{table*}
\center
\caption{The 9-planet solution of HD 10180 radial velocities. MAP estimates of the parameters and their 99\% BCS's.\label{parameters9}}
\begin{tabular}{lccc}
\hline \hline
Parameter & HD 10180 b & HD 10180 c & HD 10180 i \\
\hline
$P$ [days] & 1.17766 [1.17744, 1.17787] & 5.75973 [5.75890, 5.76047] & 9.655 [9.583,9.677] \\
$e$ & 0.05 [0.00, 0.54] & 0.07 [0.00, 0.16] & 0.05 [0.00, 0.28] \\
$K$ [ms$^{-1}$] & 0.78 [0.34, 1.21] & 4.50 [4.07, 4.92] & 0.53 [0.02, 0.99] \\
$\omega$ [rad] & 0.7 [-] & 5.8 [-] & 2.4 [] \\
$M_{0}$ [rad]& 1.2 [-] & 4.7 [-] & 1.6 [-] \\
$m_{p} \sin i$ [M$_{\oplus}$] & 1.3 [0.5, 2.1] & 13.0 [11.2, 15.0] & 1.9 [0.1, 3.5] \\
$a$ [AU] & 0.0222 [0.0211, 0.0233] & 0.0641 [0.0610, 0.0671] & 0.0904 [0.0857, 0.0947] \\
\hline
& HD 10180 d & HD 10180 e & HD 10180 j \\
\hline
$P$ [days] & 16.354 [16.341, 16.366] & 49.75 [49.68, 49.82] & 67.55 [66.67, 68.23] \\
$e$ & 0.11 [0.00, 0.24] & 0.01 [0.00, 0.11] & 0.07 [0.00, 0.19] \\
$K$ [ms$^{-1}$] & 2.90 [2.45, 3.34] & 4.14 [3.67, 4.66] & 0.75 [0.29, 1.21] \\
$\omega$ [rad] & 5.6 [-] & 1.6 [-] & 1.6 [-] \\
$M_{0}$ [rad]& 6.0 [-] & 2.0 [-] & 6.0 [-] \\
$m_{p} \sin i$ [M$_{\oplus}$] & 11.9 [9.9, 14.2] & 25.0 [21.1, 28.9] & 5.1 [1.9, 8.2] \\
$a$ [AU] & 0.1284 [0.1223, 0.1346] & 0.270 [0.257, 0.283] & 0.330 [0.314, 0.347] \\
\hline
& HD 10180 f & HD 10180 g & HD 10180 h \\
\hline
$P$ [days] & 122.88 [122.28, 123.53] & 596 [559, 626] & 2300 [2010, 2850] \\
$e$ & 0.13 [0.00, 0.28] & 0.03 [0.00, 0.43] & 0.18 [0.00, 0.44] \\
$K$ [ms$^{-1}$] & 2.86 [2.33, 3.38] & 1.61 [1.00, 2.23] & 3.02 [2.47, 3.64] \\
$\omega$ [rad] & 5.8 [-] & 0.3 [-] & 4.1 [-] \\
$M_{0}$ [rad]& 4.9 [-] & 2.6 [-] & 4.1 [-] \\
$m_{p} \sin i$ [M$_{\oplus}$] & 22.8 [18.6, 27.9] & 22.0 [13.3, 30.8] & 65.8 [52.9, 78.7] \\
$a$ [AU] & 0.494 [0.468, 0.517] & 1.415 [1.324, 1.498] & 3.49 [3.13, 4.09] \\
\hline
$\gamma$ [ms$^{-1}$] & -0.34 [-0.75, -0.01] \\
$\sigma_{j}$ [ms$^{-1}$] & 1.15 [0.92, 1.42] \\
\hline \hline
\end{tabular}
\end{table*}

Because the simple estimates in Table \ref{parameters9} can be very misleading in practice, especially in the case of nonlinear correlations between the parameters, we also plotted the projected distributions of some of the parameters in the Appendix. The distributions of $P_{i}$, $K_{i}$, and $e_{i}$, for each planet $i = 1, ..., 9$ (with indice 1 (9) referring to the shortest (longest) period), are shown in Fig. \ref{densities} and show that the periods of all Keplerian signals are indeed well constrained, the radial velocity amplitudes differ significantly from zero, and the orbital eccentricities peak at or close to zero indicating likely circular orbits. We also plotted the parameter describing the magnitude of the excess noise, or jitter, in Fig \ref{densities} to demonstrate that the MAP estimate of this parameter of 1.15 ms$^{-1}$, with a BCS of [0.92, 1.42] ms$^{-1}$, is consistent with the estimate of \citet{lovis2011} of 1.0 ms$^{-1}$, while this is not the case for the 6-Keplerian model for which the $\sigma_{j}$ receives a MAP estimate of 1.40 ms$^{-1}$ with a BCS of [1.18, 1.70] ms$^{-1}$ (Table \ref{parameters6}).

The periods of all companions get constrained well but that of the 9.66 days signal remains bimodal with two peaks at 9.66 and 9.59 days (Fig. \ref{densities}). The 9.59 days peak was found to have lower probability based on our posterior samplings because the Markov chain had good mixing properties in the sense that it visited both maxima several times during all the samplings, the posterior density had always a global maximum at 9.66 days.

\subsection{Dynamically allowed orbits}

We perform tests of dynamical stability within the context of Lagrange stability to see whether the two additional signals in the HD 10180 radial velocities could correspond to low-mass planets orbiting the star. We use the analytical approximated Lagrange stability criterion of \citet{barnes2006} to test the stability of each subsequent pairs of planets in the system. While this analytical criterion is only a rough approximation and only applicable for two planet systems, it can nevertheless provide useful information on the likely stability or instability of the system. According to \citet{barnes2006}, the orbits of two planets (denoted using subindices 1 and 2, respectively) satisfy approximately the Lagrange stability criterion if
\begin{equation}\label{stability}
  \alpha^{-3} \bigg( \mu_{1} - \frac{\mu_{2}}{\delta^{2}} \bigg) \big( \mu_{1} \gamma_{1} + \mu_{2} \gamma_{2} \delta \big)^{2} > 1 + \mu_{1}\mu_{2} \bigg( \frac{3}{\alpha} \bigg)^{4/3} ,
\end{equation} 
where $\mu_{i} = m_{i} M^{-1}$, $\alpha = \mu_{1} + \mu_{2}$, $\gamma_{i} = \sqrt{1 - e^{2}_{i}}$, $\delta = \sqrt{a_{2}/a_{1}}$, $M = m_{\star} + m_{1} + m_{2}$, $e_{i}$ is the eccentricity, $a_{i}$ is the semimajor axis, $m_{i}$ is the planetary mass, and $m_{\star}$ is stellar mass.

Using the above relation, we calculate the threshold curves for the $i$th planet with both, the next planet inside its orbit ($i-1$th) and the next planet outside its orbit ($i+1$th). We use the MAP parameter estimates for the $i$th planet and calculate the allowed eccentricities of the $i-1$th and $i+1$th planets as a function of their semimajor axes by using the MAP estimates for their masses.

\begin{figure}
\center
\includegraphics[angle=-90, width=0.49\textwidth]{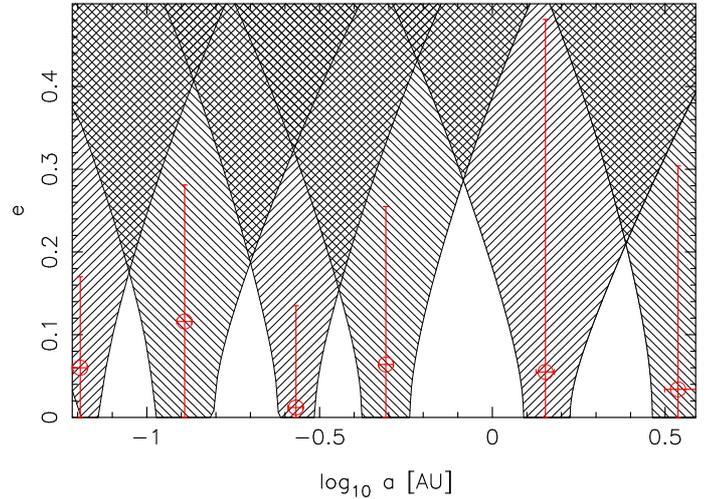}
\caption{The approximated Lagrange stability thresholds between each two planets and the MAP orbital parameters of the 6 planet solution (Table \ref{parameters6}).}\label{stability_threshold6}
\end{figure}

As \citet{lovis2011} found their 6 planet solution stable, we use it as a test case when calculating the Lagrange stability threshold curves. We use the 6 planet solution in Table \ref{parameters6}, and plot the threshold curves together with the orbital parameters in Fig. \ref{stability_threshold6}. In this Fig., the shaded areas indicate the likely unstable parameter space and the red circles indicate the positions of the modelled planets in the system.

It can be seen in Fig. \ref{stability_threshold6}, that the $i$th planet has orbital parameters that keeps it inside the Lagrange stability region of the neighbouring planetary companions for all $i = 1, ..., 6$. This result then agrees with the numerical integrations of \citet{lovis2011} and, while only a rough approximation of the reality, encourages us to use the criterion in Eq. (\ref{stability}) for our 7-,  8-, and 9-companion solutions as well. Fig. \ref{stability_threshold6} also suggests, that there might be stable regions between the orbits of these 6 planets for additional low-mass companions.



\begin{figure}
\center
\includegraphics[angle=-90, width=0.49\textwidth]{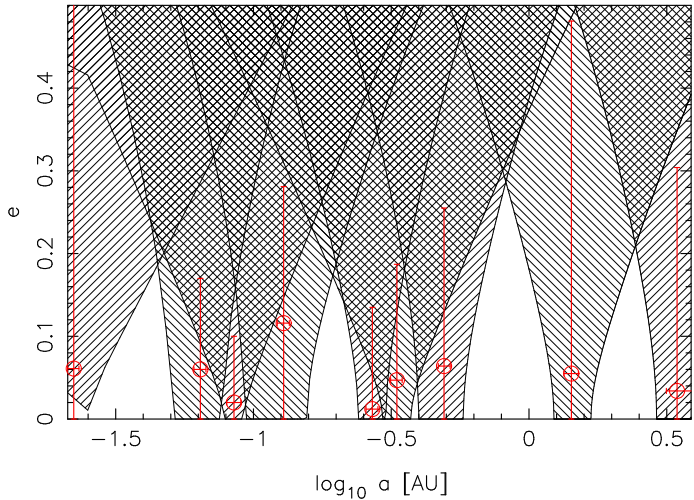}
\caption{The approximated Lagrange stability thresholds between each two planets and the MAP orbital parameters of the 9 planet solution (Table \ref{parameters9}).}\label{stability_threshold9}
\end{figure}

When interpreting all the signals in our 9-planet solution as being of planetary origin, the stability thresholds show some interesting features (Fig. \ref{stability_threshold9}). The periodicities at 9.66 and 67.6 days would correspond to planets that satisfy the condition in Eq. (\ref{stability}) if the orbital eccentricities of all the companions were close to or below the MAP estimate, which, according to the probability densities in Fig. \ref{densities}, appears to be likely based on the data alone. This means that the planetary origin of these periodicities cannot be ruled out by this analysis.

In reality, the stability constraints are necessarily more limiting than those described by the simple Eq. \ref{stability} because of the gravitational interactions between all the planets, not only the nearby ones. Also, it does not take the stabilising or destabilising effects of mean motion resonances. However, the numerical integrations of the orbits of the seven planets performed by \citet{lovis2011}, actually do not rule the 0.09 and 0.33 AU orbits (Table \ref{parameters9}) out as unstable but show that there are regions of at least ''reasonable stability`` in the vicinity of these orbits given that they are close-circular and that the planetary masses in these orbits are small. When interpreted as being of planetary origin, the periodic signals at roughly 9.66 and 67.6 days satisfy these requirements.

We note, that the selected prior density for the orbital eccentricities, namely $\pi(e_{i}) \propto \mathcal{N}(0, \sigma_{e}^{2})$ for all $i$, in fact helps slightly in removing \emph{a priori} unstable solutions from the parameter posterior density. However, this effect is not very significant in this case, because the prior requirement that the signals in the data do not correspond to planets with crossing orbits constrains the eccentricities much more strongly and the parameter posteriors we would receive with uniform eccentricity prior would therefore not differ significantly from those reported in Table \ref{parameters9} and Figs. \ref{orbits} and \ref{densities}.

\subsection{Avoiding unconstrained solutions}

To further emphasise our confidence in the 9 periodic signals in the data, we tried finding additional signals in the gaps between the 9-Keplerians, especially, between the 123 and 600 days orbits. This part of the period-space is interesting because any habitable planet in the system would have its orbital period in this space and because the stability thresholds allow the existence of low-mass planets in close-circular orbits in this region (see Fig. \ref{stability_threshold9}).

\begin{figure}
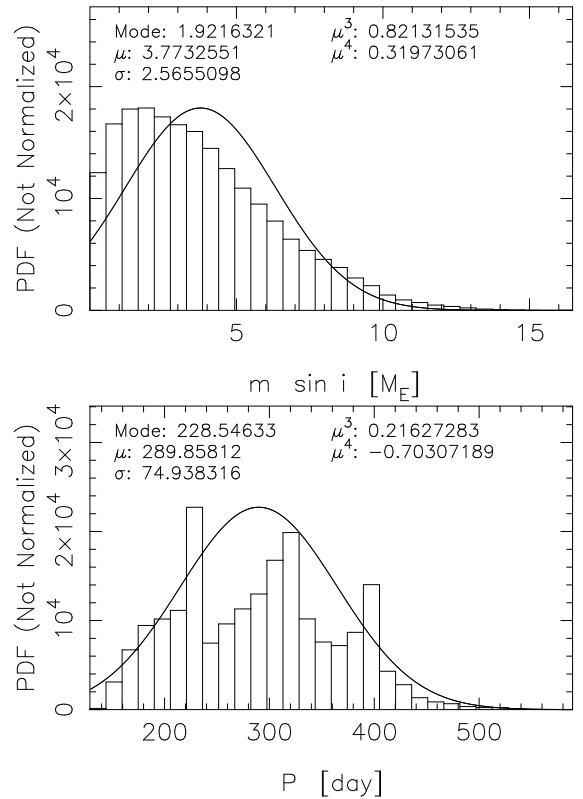

\center
\includegraphics[angle=-90,width=0.40\textwidth]{rvdist10_rv_HD10180_dist_m.ps}
\includegraphics[angle=-90,width=0.40\textwidth]{rvdist10_rv_HD10180_dist_P.ps}
\caption{The posterior densities of the minimum mass and orbital period of the planet that could exist in the habitable zone of HD 10180 without having been detected using the current data. The solid curve is a Gaussian density with the same mean ($\mu$) and variance ($\sigma^{2}$) as the parameter distribution has.}\label{mass_limit}
\end{figure}

As could already be suspected based on the periodograms of residuals in Fig. \ref{periodograms}, we could not find any signals between the periods of 123 and 600 days. The sampling of the parameter space of this 10-Keplerian model was much more difficult than that of the models with fewer signals because the orbital period of this hypothetical 10th signal was only constrained by the fact that \emph{a priori} we did not allow orbital crossings (Fig. \ref{mass_limit}). As a result, the probability density of the orbital period did not have a clear maximum but several small maxima, whose relative significance is not known because we cannot be sure whether the Markov chain converged to the posterior in this case (Fig. \ref{mass_limit}). Therefore, the distribution of the orbital period in Fig. \ref{mass_limit} is only a rough estimate of what the density might look like.

Because different samplings yielded similar but not equal densities for the orbital period, we could not be sure whether the chain had indeed converged to the posterior or not. For this reason, we did not consider the corresponding posterior probability of this model trustworthy and do not show it in Table \ref{model_probabilities}. The posterior probabilities we received were roughly 1-5\% of that of the 9-Keplerian model. However, these samplings still provide some interesting information in the sense that we can put an upper limit to the planetary masses that could exist between the 123 and 600 days periods and still not be detected confidently by the current observations.

According to the samplings of the parameter space, the probability of there being Keplerian signals between 123 and 600 days periods with radial velocity amplitudes in excess of 1.1 ms$^{-1}$ is less than 1\%. Our MAP solution for this amplitude is 0.12 ms$^{-1}$ with a BCS of [0.00, 1.10] ms$^{-1}$. This means that we can rule out the existence of planets more massive than approximately 12.1 M$_{\oplus}$ in this period space because such companions could have been detected by the current data. Therefore, even the fact that a signal was not detected can help constraining the properties of the system, as seen in the probability density of the minimum mass of this hypothetical planet in Fig. \ref{mass_limit} -- clearly this signal is indistinguishable from one with negligible amplitude, as the density is peaking close to zero. This result means that an Earth mass planet could exist in the habitable zone of HD 10180 and most likely, if there is a low-mass companion orbiting the star between orbital periods of 123 and 600 days, it has a minimum mass of less than 12.1 M$_{\oplus}$. However, we cannot say much about the possible orbit of this hypothetical companion, because all the orbits between 123 and 600 days without orbital crossings are almost equally probable (Fig. \ref{mass_limit}). All we can say, is that orbital crossings limit the allowed periodicities and yield a 99\% BCS of [128, 534] days for this orbital period, though, further dynamical constraints would narrow this interval even further, as shown in Fig. \ref{stability_threshold9} and Fig. 12 of \citet{lovis2011}.

\subsection{Detectability of Keplerian signals}

To further emphasise the significance of the 9 signals we detect in the HD 10180 radial velocities, we generated an artificial data set to see if known signals could be extracted from it confidently given the definition of our criteria for detection threshold. This data set was generated by using the same 190 epochs as in the HD 10180 data of \citet{lovis2011}. We generated the radial velocities corresponding to these epochs by using a superposition of the 9 signals with parameters roughly as in Table \ref{parameters9}. Further, we added three noise components, Gaussian noise with zero mean as described by the uncertainty estimates of each original radial velocity of \citet{lovis2011}, Gaussian noise with zero mean and $\sigma = 1.1$ ms$^{-1}$ and uniform noise as a random number drawn from the interval [-0.2, 0.2] ms$^{-1}$. The lattter two produce together the observed excess noise in the data of roughly 1.15 ms$^{-1}$ when modelled as pure Gaussian noise. We used this different noise model when generating the data to not commit an inverse crime, i.e. to not generate the data using the same model used to analyse it which would correspond to studying the properties of the model only \citep[see e.g.][]{kaipio2005}.

\begin{figure}
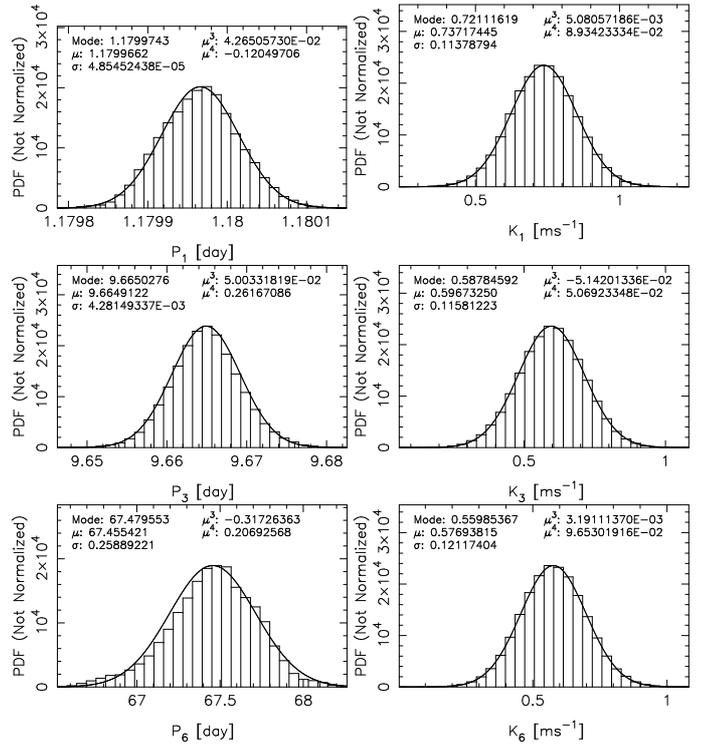

\center
\includegraphics[angle=-90,width=0.24\textwidth]{rvdist9_rv_HD10180a_dist_P1.ps}
\includegraphics[angle=-90,width=0.24\textwidth]{rvdist9_rv_HD10180a_dist_K1.ps}

\includegraphics[angle=-90,width=0.24\textwidth]{rvdist9_rv_HD10180a_dist_P3.ps}
\includegraphics[angle=-90,width=0.24\textwidth]{rvdist9_rv_HD10180a_dist_K3.ps}

\includegraphics[angle=-90,width=0.24\textwidth]{rvdist9_rv_HD10180a_dist_P6.ps}
\includegraphics[angle=-90,width=0.24\textwidth]{rvdist9_rv_HD10180a_dist_K6.ps}
\caption{The distributions of the periods and amplitudes received for the three weakest signals in the artificially generated data. The added signals had $P_{i} =$ 1.18, 9.66, and 67.55 days, and $K_{i} =$ 0.78, 0.53, and 0.75 ms$^{-1}$ for $i = 1, 3, 6$, respectively.}\label{test_case}
\end{figure}

Analysing this artificial data yielded results confirming the trustworthiness of our methods. Using the detection criteria defined above, we could extract all 9 signals from the artificial data with well constrained amplitudes and periods that were consistent with the added signals in the sense that the 99\% BCS's of the parameters contained the values of the added signals. The MAP estimates did differ from the added signals but not significantly so given the uncertainties as described by distributions corresponding to the parameter posterior density. This simply represents the statistical nature of the solutions based on posterior samplings. As an example, we show the periods and amplitudes of the three weakest signals with the lowes radial velocity amplitudes as probability distributions (Fig. \ref{test_case}). This Fig. indicates that if the radial velocity noise is indeed dominated by Gaussian noise, the low-amplitude signals we report can be detected confidently. This conclusion was also supported by the corresponding model probabilities we received for the artificial data set. These probabilities indicated that the 9-Keplerian model had the greatest posterior probability exceedind the threshold of 150 times greater than the probability of the 8-Keplerian model.

\subsection{Comparison with earlier results}

The analyses of \citet{lovis2011} of the same radial velocities yielded differing results, i.e. the number of significant powers in periodogram was found to be 7 instead of the 9 significant periodic signals reported here. While this difference is likely due to the fact that the power spectrums are calculated by fixing the parameters of the previous signals to some point estimates when searching for additional peaks, the analyses of \citet{feroz2011} suffer from similar sources of bias. The Bayesian approach of \citet{feroz2011} was basically a search of $k+1$th signal in the residuals of the model with $k$ Keplerian signals. They derived the probability densities of the residuals by assuming they were uncorrelated, which is unlikely to be the case, especially, if there are signals left in the residuals. Therefore, any significance test, in this case the comparison of Bayesian evidences of a null hypothesis and an alternative one with a model containing one more signal, is similarly biased by the fact that these correlations are not fully accounted for. While this source of bias might be relatively small, in the approach of \citet{feroz2011}, the effective number of parameters is artificially decreased by the very fact that residuals are being analysed. This decrease, in turn, might make any comparisons of Bayesian evidences biased. Modifying the uncertainties corresponding to the posterior density of the model residuals does not necessarily account for this decrease in the dimension when the weakest signals among the $k$ detected ones are at or below the residual uncertainty and their contribution to the total uncertainty of the residuals becomes negligible in the first place.

In fact, we could replicate the results of \citet{feroz2011} using the OBMH estimates of marginal integrals.  Using their simple method, we received a result that the 6-Keplerian model residuals could not be found to contain one more signal. While the Bayesian evidence for this additional signal did not exceed that of having the 6-Keplerian model residuals consist of purely Gaussian noise, the amplitude of this additional signal in the residuals was also found indistinguishable from zero, in accordance with our criteria of not detecting a signal. Therefore, the results we present here do not actually conflict with those of \citet{feroz2011}. As an example, the log-Bayesian evidences of 6 and 7-Keplerian models (Table \ref{model_probabilities}) were -358.36 and -343.73, respectively. Analysing the residuals of the model with $k=6$ using 0- and 1-Keplerian models should yield the similar numbers (up to differences in the priors), if the method of \citet{feroz2011} were trustworthy. Instead, we received values of roughly -338 for both models. This means, that the null-hypothesis is exaggerated because of the fact that the corresponding model contains only two free parameters, the jitter magnitude and the reference velocity. The Occamian principle cannot therefore penalise this model as much as it should, and the results are biased in favour of the null-hypothesis, which effectively prevents the detection of low-amplitude signals.

We also tested the method of \citet{feroz2011} in analysing the artificial test data described in the previous subsection. The results were almost similar: the log-Bayesian evidences given the residuals of a model with $k$ Keplerian signals were found to favour the 6-Keplerian interpretation, while the artificial data was known to contain 9 signals. This further emphasises the fact, that the method of \citet{feroz2011}, while capable of detecting the strongest signals in the data, cannot be considered trustworthy if it fails to make a positive detection of a low-amplitude signal. Yet, it is likely trustworthy if it provides a positive detection.

\section{Conclusions and Discussion}

We have re-analysed the 190 HARPS radial velocities of HD 10180 published in \citet{lovis2011} and report our findings in this article. First, we have revised the orbital parameters of the proposed 6 planetary companions to this star and calculated realistic uncertainty estimates based on samplings of the parameter space. We also verified the significance of the 1.18 days signal reported by \citet{lovis2011} and interpreted as arising from a planetary companion with this orbital period and a minimum mass of as low as 1.3 M$_{\oplus}$. In addition to these seven signals, we report two additional periodic signals that are, according to our model probabilities in Table \ref{model_probabilities}, statistically significant and unlikely to be caused by noise or data sampling or poor phase-coverage of the observations. Their amplitudes are well constrained and differ statistically from zero, which would not be the case unless they corresponded to actual periodicities in the data. We can also constrain their periods from above and below reasonably accurately.

A related analysis of the same radial velocities was recently carried out by \citet{feroz2011} but they received results differing in the number of significant periodicities. They claimed that only 6 signals can be detected reliably in the data, as opposed to 9 detected in the current work. However, they first analysed the data using a model with $k$ Keplerian signals and analysed the remaining residuals to see if they contained one more, $k+1$th signal. While, as in the analyses of \citet{lovis2011}, this approach does not fully account for the uncertainties of the first $k$ signals when they have low amplitudes and their contribution to the residual uncertainty is negligible, it also actually assumes that the $k$-Keplerian model is a correct one and then tests if it is not so, which is a clear contradiction and, while useful in case of strong signals, as demonstrated by \citet{feroz2011}, likely prevents the detection of weak signals in the data. This is underlined by the fact that \citet{feroz2011} assume the residual vector to have an uncorrelated multivariate Gaussian distribution -- this clearly cannot be the case if there are signals left in the residuals. Our analyses are not prone to similar weaknesses.

Because planetary companions orbiting the star would produce the kind of periodicities we observe in the radial velocities, the interpretation of the two new signals as two new low-mass planets seems reasonable. As noted by \citet{lovis2011}, the star is a very quiet one without clear activity-induced periodicities, which makes it unlikely that one or some of the periodic signals in the data were caused by stellar phenomena. Also, the periodicities we report, namely 9.66 and 67.6 days, do not coincide with any periodicities arising from the movement of the bodies in the Solar system. Therefore, we consider the interpretation of these two new signals of being of planetary origin to be the most credible explanation. If this was the case, these two signals would correspond to planets on close-circular orbits with minimum masses of 1.9$^{+1.6}_{-1.8}$ and 5.1$^{+3.1}_{-3.2}$ M$_{\oplus}$, respectively, enabling the classification of them as super-Earths.

Apart from the significance of the signals we observe, there is another rather strong argument in favour of the interpretation that all nine signals in the data are actually of planetary origin. Assuming that they were not, which based on stability reasons is the case with the 6.51 days signal that is quite certainly an alias of the 1.18 days periodicity likely caused by a planet \citep{lovis2011}, we would expect the weakest signals to be at random periods independent of the six strong periodicities in the data and the seventh 1.18 periodicity. Instead, this is not the case but the two additional signals reported in the current work appear at periods that fall in between the existing ones and, if interpreted as being of planetary origin, likely have orbits that enable long-term stability of the system (Fig. \ref{stability_threshold9}) if their orbital eccentricities are close to or below the estimates in Table \ref{parameters9}. As stated by \citet{lovis2011}, there are ''empty`` places in the HD 10180 system that allow dynamical stability of low-mass planets in the orbits corresponding to these empty places in the orbital parameter space, especially in the $a-e$ space. The two periodic signals we observe in the data correspond exactly to those empty places if interpreted as being of planetary origin.

Additional measurements are needed to verify the significance of the two new periodic signals in the radial velocities of HD 10180 and to set tighter constraints to the orbital parameters of the planets in the system. Also, the possibility that all nine signals in the data correspond to planets should be tested by full-scale numerical integrations of their orbits. If all the configurations allowed by our solution in Table \ref{parameters9} were found to correspond to unstable systems in any timescales of less than the estimated stellar age of 4.3$\pm$0.5 Gyr \citep{lovis2011}, it would be a strong argument against the planetary interpretation of one or both of the signals we report in this article or the third low-amplitude 1.18 days signal. However, the results we present here and those in \citet{lovis2011} suggest that this planetary interpretation of all the signals cannot be ruled out by dynamical analyses of the system.

If the significance of these signals increases when additional high-precision radial velocities become available, and their interpretation as being of planetary origin is confirmed, the planetary system aroung HD 10180 will be the first one to top the Solar system in terms of number of planets in its orbits. Further, according to the rough dynamical considerations of the current work and the more extensive numerical integrations of \citet{lovis2011}, there are stable orbits for a low-mass companion in or around the habitable zone of the star. If such a companion exists, its minimum mass is unlikely to exceed 12.1 M$_{\oplus}$ according to our posterior samplings of the corresponding parameter space.

\begin{acknowledgements}
M. Tuomi is supported by RoPACS (Rocky Planets Around Cools Stars), a Marie Curie Initial Training Network funded by the European Commission's Seventh Framework Programme. The author would like to thank the two anonymous referees for constructive comments and suggestions that resulted in significant improvements in the article.
\end{acknowledgements}

\appendix

\section{Parameter distributions}

\begin{figure*}
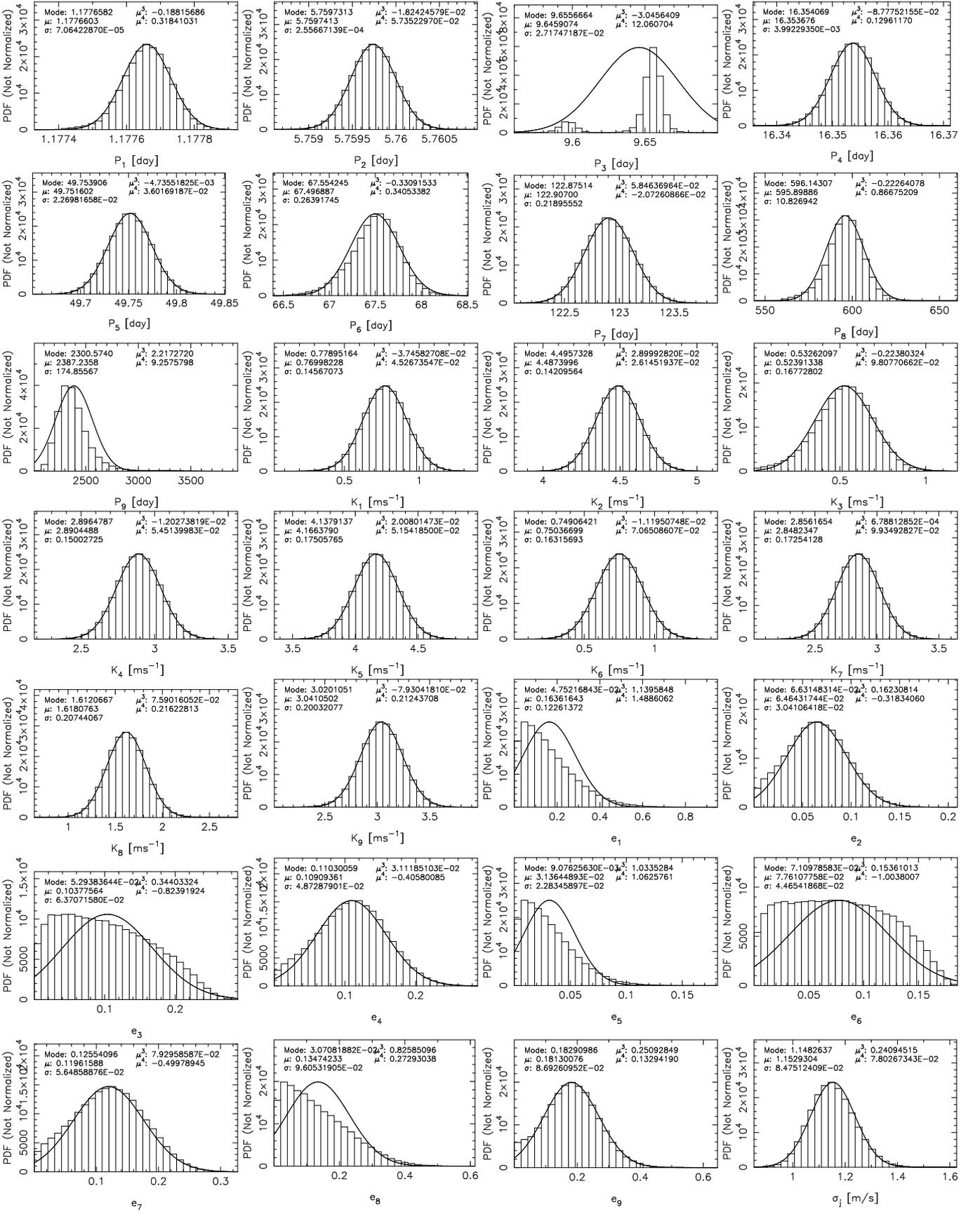

\center
\includegraphics[angle=-90,width=0.24\textwidth]{rvdist9_rv_HD10180_dist_P1.ps}
\includegraphics[angle=-90,width=0.24\textwidth]{rvdist9_rv_HD10180_dist_P2.ps}
\includegraphics[angle=-90,width=0.24\textwidth]{rvdist9_rv_HD10180_dist_P3.ps}
\includegraphics[angle=-90,width=0.24\textwidth]{rvdist9_rv_HD10180_dist_P4.ps}

\includegraphics[angle=-90,width=0.24\textwidth]{rvdist9_rv_HD10180_dist_P5.ps}
\includegraphics[angle=-90,width=0.24\textwidth]{rvdist9_rv_HD10180_dist_P6.ps}
\includegraphics[angle=-90,width=0.24\textwidth]{rvdist9_rv_HD10180_dist_P7.ps}
\includegraphics[angle=-90,width=0.24\textwidth]{rvdist9_rv_HD10180_dist_P8.ps}

\includegraphics[angle=-90,width=0.24\textwidth]{rvdist9_rv_HD10180_dist_P9.ps}
\includegraphics[angle=-90,width=0.24\textwidth]{rvdist9_rv_HD10180_dist_K1.ps}
\includegraphics[angle=-90,width=0.24\textwidth]{rvdist9_rv_HD10180_dist_K2.ps}
\includegraphics[angle=-90,width=0.24\textwidth]{rvdist9_rv_HD10180_dist_K3.ps}

\includegraphics[angle=-90,width=0.24\textwidth]{rvdist9_rv_HD10180_dist_K4.ps}
\includegraphics[angle=-90,width=0.24\textwidth]{rvdist9_rv_HD10180_dist_K5.ps}
\includegraphics[angle=-90,width=0.24\textwidth]{rvdist9_rv_HD10180_dist_K6.ps}
\includegraphics[angle=-90,width=0.24\textwidth]{rvdist9_rv_HD10180_dist_K7.ps}

\includegraphics[angle=-90,width=0.24\textwidth]{rvdist9_rv_HD10180_dist_K8.ps}
\includegraphics[angle=-90,width=0.24\textwidth]{rvdist9_rv_HD10180_dist_K9.ps}
\includegraphics[angle=-90,width=0.24\textwidth]{rvdist9_rv_HD10180_dist_e1.ps}
\includegraphics[angle=-90,width=0.24\textwidth]{rvdist9_rv_HD10180_dist_e2.ps}

\includegraphics[angle=-90,width=0.24\textwidth]{rvdist9_rv_HD10180_dist_e3.ps}
\includegraphics[angle=-90,width=0.24\textwidth]{rvdist9_rv_HD10180_dist_e4.ps}
\includegraphics[angle=-90,width=0.24\textwidth]{rvdist9_rv_HD10180_dist_e5.ps}
\includegraphics[angle=-90,width=0.24\textwidth]{rvdist9_rv_HD10180_dist_e6.ps}

\includegraphics[angle=-90,width=0.24\textwidth]{rvdist9_rv_HD10180_dist_e7.ps}
\includegraphics[angle=-90,width=0.24\textwidth]{rvdist9_rv_HD10180_dist_e8.ps}
\includegraphics[angle=-90,width=0.24\textwidth]{rvdist9_rv_HD10180_dist_e9.ps}
\includegraphics[angle=-90,width=0.24\textwidth]{rvdist9_rv_HD10180_dist_sj.ps}
\caption{The posterior densities of orbital periods ($P_{i}$), eccentricities ($e_{i}$), and radial velocity amplitudes ($K_{i}$) corresponding to the 9 Keplerian signals in the data and of the magnitude of the excess jitter ($ \sigma_{j}$). The solid curve is a Gaussian density with the same mean ($\mu$) and variance ($\sigma^{2}$) as the parameter distribution has.}\label{densities}
\end{figure*}



\begin{thebibliography}{100}\small
\bibitem[\protect\astroncite{Barnes \& Greenberg}{2006}]{barnes2006} Barnes, R. \& Greenberg, R. 2006, ApJ, 647, L163
\bibitem[\protect\astroncite{Chib \& Jeliazkov}{2001}]{chib2001} Chib S. \& Jeliazkov I. 2001, J. Am. Stat. Ass., 96, 270
\bibitem[\protect\astroncite{Clyde et al.}{2007}]{clyde2007} Clyde, M. A., Berger, J. O., Bullard, F., et al. 2007, Statistical Challenges in Modern Astronomy IV, Babu, G. J. \& Feigelson, E. D. (eds.), ASP Conf. Ser., 371, 224
\bibitem[\protect\astroncite{Feroz et al.}{2011}]{feroz2011} Feroz, F., Balan, S. T., \& Hobson, M. P. 2011, MNRAS, 415, 3462
\bibitem[\protect\astroncite{Ford \& Gregory}{2007}]{ford2007} Ford, E. B. \& Gregory, P. C. 2007, Statistical Challenges in Modern Astronomy IV, Babu, G. J. \& Feigelson, E. D. (eds.), ASP Conf. Ser., 371, 189
\bibitem[\protect\astroncite{Gelman et al.}{1996}]{gelman1996} Gelman, A. G., Roberts, G. O., \& Gilks, W. R. 1996, Efficient Metropolis jumping rules. In Bernardo, J. M., Berger, J. O., David, A. F., \& Smith, A. F. M. (eds.), Bayesian Statistics V, pp. 599
\bibitem[\protect\astroncite{Gregory}{2005}]{gregory2005} Gregory, P. C. 2005, ApJ, 631, 1198
\bibitem[\protect\astroncite{Gregory}{2007a}]{gregory2007a} Gregory, P. C. 2007a, MNRAS, 381, 1607
\bibitem[\protect\astroncite{Gregory}{2007b}]{gregory2007b} Gregory, P. C. 2007b, MNRAS, 374, 1321
\bibitem[\protect\astroncite{Gregory}{2011}]{gregory2011} Gregory, P. C. 2011, MNRAS, 415, 2523
\bibitem[\protect\astroncite{Haario et al.}{2001}]{haario2001} Haario, H., Saksman, E., \& Tamminen, J. 2001, Bernoulli, 7, 223
\bibitem[\protect\astroncite{Hastings}{1970}]{hastings1970} Hastings, W. 1970, Biometrika 57, 97
\bibitem[\protect\astroncite{Kaipio \& Somersalo}{2005}]{kaipio2005} Kaipio, J. \& Somersalo, E. 2005. Statistical and Computational Inverse Problems, Applied Mathematical Sciences 160 (Springer)
\bibitem[\protect\astroncite{Kass \& Raftery}{1995}]{kass1995} Kass, R. E. \& Raftery, A. E. 1995, J. Am. Stat. Ass., 430, 773
\bibitem[\protect\astroncite{Lissauer et al.}{2011}]{lissauer2011} Lissauer, J. J., Fabrycky, D. C., Ford, E. B., et al. 2011, Nature, 470, 53
\bibitem[\protect\astroncite{Lomb}{1976}]{lomb1976} Lomb, N. R. 1976, Astrophys. Space Sci., 39, 447
\bibitem[\protect\astroncite{Loredo et al.}{2011}]{loredo2011} Loredo, T. J., Berger, J. O., Chernoff, D. F., et al. 2011, Statistical Methodology, accepted (doi:10.1016/j.stamet.2011.07.005)
\bibitem[\protect\astroncite{Lovis et al.}{2006}]{lovis2006} Lovis, C., Mayor, M., Pepe, F., et al. 2006, Nature, 441, 305
\bibitem[\protect\astroncite{Lovis et al.}{2011}]{lovis2011} Lovis, C., S\'egransan, D., Mayor, M., et al. 2011, A\&A, 528, A112
\bibitem[\protect\astroncite{Mayor et al.}{2003}]{mayor2003} Mayor, M., Pepe, F., Queloz, D., et al. 2003, Messenger, 114, 20
\bibitem[\protect\astroncite{Mayor et al.}{2009}]{mayor2009} Mayor, M., Bonfils, X., Forveille, T., et al. 2009, A\&A, 507, 487
\bibitem[\protect\astroncite{Mayor et al.}{2011}]{mayor2011} Mayor, M., Marmier, M., Lovis, C., et al. 2011, A\&A, accepted (arXiv:1109.2497 [astro-ph.EP])
\bibitem[\protect\astroncite{Metropolis et al.}{1953}]{metropolis1953} Metropolis, N., Rosenbluth, A. W., Rosenbluth, M. N., et al. 1953, J. Chem. Phys., 21, 1087
\bibitem[\protect\astroncite{Pepe et al.}{2011}]{pepe2011} Pepe, F., Lovis, C., S\'egransan, D., et al 2011, A\&A, 534, A58
\bibitem[\protect\astroncite{Scargle}{1982}]{scargle1982} Scargle, J. D. 1982, ApJ, 263, 835
\bibitem[\protect\astroncite{Tuomi \& Kotiranta}{2009}]{tuomi2009} Tuomi, M. \& Kotiranta, S. 2009, A\&A, 496, L13
\bibitem[\protect\astroncite{Tuomi}{2011}]{tuomi2011} Tuomi, M. 2011, A\&A, 528, L5
\bibitem[\protect\astroncite{Tuomi et al.}{2009}]{tuomi2009a} Tuomi, M., Kotiranta, S., \& Kaasalainen, M. 2009, A\&A, 494, 769
\bibitem[\protect\astroncite{Tuomi et al.}{2011}]{tuomi2011b} Tuomi, M., Pinfield, D., \& Jones, H. R. A. 2011, A\&A, 532, A116
\bibitem[\protect\astroncite{von Paris et al.}{2011}]{vonparis2011} von Paris, P., Gebauer, S., Godolt, M., et al. 2011, A\&A, 532, A58
\bibitem[\protect\astroncite{Vogt et al.}{2010}]{vogt2010} Vogt, S. S., Butler, R. P., Rivera, E. J., et al. 2010, ApJ, 723, 954
\bibitem[\protect\astroncite{Wordsworth et al.}{2010}]{wordsworth2010} Wordsworth, R., Forget, F., Selsis, F., et al. 2010, A\&A, 522, A22
\end{thebibliography}
\end{document}